\documentclass[12pt,titlepage]{article}
\usepackage[english]{babel}
\usepackage{amsmath,a4}
\usepackage{epsfig}
\begin{document}


\begin{titlepage}
\begin{flushright}
\end{flushright}

\vspace{.5cm}
\begin{center}
{\huge \bf Effects of Littlest Higgs model}

\vspace{0.3cm}

{\huge \bf  in rare D meson decays}

\vspace{.7cm}

{\large  Svjetlana Fajfer and Sasa Prelovsek \\}

\vspace{0.5cm}

{\it Department of Physics, University of Ljubljana, 
Jadranska 19, 1000 Ljubljana, Slovenia}

\vspace{0.1cm}

and

\vspace{0.1cm}

{\it J. Stefan Institute, Jamova 39, P. O. Box 300, 1001 Ljubljana, 
Slovenia}

\vspace{.5cm}

\end{center}
\centerline{\large \bf ABSTRACT}

\vspace{0.5cm}
A tree-level flavor changing neutral current in the up-like quark sector  
appears in one of the variations of the Littlest Higgs model. 
We investigate the effects of this coupling  
in the $D^+ \to \pi^+ l^+ l^-$  and
$D^0 \to \rho^0 l^+ l^-$ decays, which are the  most appropriate 
candidates for the experimental studies.  However, the effects 
are found to be 
too small to be observed in the current and the 
foreseen experimental facilities. 
These decays are still 
dominated by the  standard model long-distance contributions, 
which are reevaluated based  on the new experimental input.  

\end{titlepage}

\section{ Introduction}
\vspace{.5cm}

The effects of new physics in  hadronic phenomena 
are most likely to be seen 
 in the down-like quark sector.  Many new scenarios modify 
the flavor changing natural currents (FCNC)
 with respect to Standard Model (SM) framework. 
This might lead to observable effects in the processes 
which do not appear  at the tree-level within SM  
 like $ b \to s $, $b \to d$, $s \to d$,
$b \bar s \leftrightarrow \bar b s$, $b \bar d \leftrightarrow \bar b d$  and 
$s \bar d \leftrightarrow \bar s d$ transitions. 
  Most of the charm meson processes,  where 
$c \to u$  and $c \bar u \leftrightarrow \bar c u$ transitions occur,  
are however dominated by the standard model long-distance  contributions
 \cite{burdman1} - \cite{bigi0}.

On the experimental side there are many studies of rare charm meson decays. 
The first  observed rare $D$ meson decay was  the radiative
 weak decay 
$D \to \phi \gamma$. Its rate $BR(D \to \phi \gamma)= 2.6^{+0.7}_{-0.6} \times 10^{-5}$ has been 
measured by Belle collaboration 
\cite{Belle1}  and hopefully other
 radiative weak charm decays will be observed soon \cite{CLEO_pll}. 
 The hadronic decays, in which the   $c \to u $ transition occur,  are interesting for 
the searches of new physics. The $c \to u \gamma$ decay rate 
is strongly GIM suppressed at the leading order in the SM, while the QCD effects 
enhance it up to the order of $10^{-8}$ \cite{greub}. The 
minimal super-symmetric standard model (MSSM) can increase this rate 
by a  factor of $100$  \cite{sasa}. 
 On the other hand, the long distance  contributions in the relevant 
$D \to V\gamma$  decays 
 ($V$ is a light vector meson) 
give the branching ratios of the order $Br\sim 10^{-6}$ 
\cite{burdman1,prelovsek1}, which makes the 
search for new physics  in radiative charm decays 
almost impossible.

Another possibility to 
search for  the effects of new physics  in the charm sector is 
offered in the studies of  
$D \to X l^+ l^-$ decays which might be results of the  $c \to u l^+ l^-$   
 FCNC transition 
\cite{burdman2,burdman3,prelovsek0,prelovsek2,prelovsek3}. Here $X$ 
is light vector meson $V$ or pseudoscalar meson $P$. 
The leading order rate for the inclusive $c\to u l^+l^-$ calculated within 
 SM \cite{prelovsek3}
was found to be suppressed by QCD corrections  \cite{burdman2}. 
The inclusion of the renormalization group equations  
for the Wilson coefficients 
gave an additional significant 
suppression  leading to the rates  
$\Gamma(c\to ue^+e^-)/\Gamma_{D^0}=2.4\times 10^{-10}$ and
$\Gamma(c\to u\mu^+\mu^-)/\Gamma_{D^0}=0.5\times 10^{-10}$ \cite{jure}.   
These transitions are largely driven by a virtual photon at low 
dilepton mass $m_{ll}\equiv \sqrt{(p_++p_-)^2}$.   
The total rate for $D \to X l^+ l^-$ is dominated by the 
long distance resonant contributions at dilepton mass 
$m_{ll}=m_\rho,m_\omega,m_\phi$ 
and even the largest contributions from new physics are not expected to 
affect the total rate significantly \cite{burdman2, prelovsek3}. 
New physics could only modify the dilepton mass distribution  
 below 
$\rho$ or distribution above $\phi$. 
 In the case of $D\to\pi l^+l^-$ there is a broad  
kinematical region of dilepton mass 
 above $\phi$ resonance which presents an unique possibility to 
study $c\to ul^+l^-$ at high $m_{ll}$ \cite{prelovsek3}. 
The leading 
contribution to $c\to ul^+l^-$ in general MSSM with the conserved R parity 
comes from one-loop diagram with 
gluino and squarks in the loop \cite{burdman2,prelovsek3,sasa}. 
It proceeds via virtual photon  
and significantly enhances the $c\to ul^+l^-$ 
spectrum at small $m_{ll}$. This MSSM enhancement is not so drastic in the 
hadronic decays, since the gauge invariance in $D\to Pl^+l^-$ 
imposes an 
additional factor of $m_{ll}^2$ \cite{burdman2,prelovsek3}, while 
 $D\to Vl^+l^-$ has large long distance contributions at small $m_{ll}$ 
just like $D\to V\gamma$. The R-parity violating SUSY contributions 
can induce $c\to ul^+l^-$ at tree level  via sparticle exchange. 
This  
can give a sizable enhancement of decay width distribution  
at low and at high 
$m_{ll}$ \cite{burdman2,burdman3}.  The presence of the 
 R-parity violating couplings modifies also the forward - backward 
asymmetry in the case of 
$D \to Vl^+ l^-$ 
decay. There are intensive experimental efforts by CLEO 
\cite{CLEO_pll,CLEO_vll} 
 and FERMILAB \cite{FERMILAB_pll,FERMILAB_vll} collaborations to improve 
the upper limits on the rates for $D\to X l^+l^-$ decays. 
Two events in the channel 
$D^+\to \pi^+e^+e^-$ with $m_{ee}$ close to $m_\phi$ have 
already been observed by CLEO \cite{CLEO_pll}. 

The other rare $D$ meson decays are not so easily accessible by experimental 
searches. The $c\to u$ transition occurs also in  
$D^0 \to l^+ l^-$ decay. However, 
in the SM this mode is 
helicity suppressed and  also dominated by the long distance contributions 
\cite{burdman2,prelovsek0,bigi0}
leading to the rate of the order $10^{-13}$. 

Among many extensions of the Standard Model, the  Littlest Higgs (LH) 
model  (see e.g. \cite{arkani} - \cite{csaki}) 
offers a simple and appealing solution to the gauge hierarchy problem. 
 The Higgs boson of this model is a pseudo Goldstone boson of a new global symmetry, 
which is spontaneously broken at the scale $4\pi f$. This protects Higgs mass 
 against quadratic divergences from self interactions. 
The quadratic divergences in the Higgs mass  due to the SM gauge bosons are 
cancelled  by the contributions 
 of the new heavy gauge bosons with spin 1. The divergence 
due to  top quark is cancelled by the 
contribution of the new heavy quark with the charge $2/3$ and spin $1/2$ . 
There are two interesting consequences that arise 
 from  the existence of this new 
 quark which is a $SU(2)_L$ singlet \cite{lee}. 
It extends the $3\times 3$ CKM matrix in SM to a $4\times 3$ matrix. 
It also allows Z-mediated FCNC at tree-level in the up sector 
but not in the down sector \cite{lee}. 
 The magnitude of the relevant tree-level $c \to u Z$ coupling  
$|V_{ub}||V_{cb}|v^2/f^2$ 
is constrained via the scale $f\geq {\cal O}(1~{\rm TeV})$ 
 by the precision electro-weak observables \cite{csaki}.  
In ref. \cite{lee} the author has studied effects of this new FCNC coupling 
in $ D \to \mu ^+ \mu^-$, $ D^0 \leftrightarrow \bar D^0$ 
oscillations and $t \to c Z$ decay. The effects were found to be 
insignificant for the current 
experimental studies and the testability of the model requires 
more stringent measurements of the mixing angles at 
Large Hadron Collider.

 In this paper we investigate possible effects 
of the tree-level $c\to uZ$ coupling from the Littlest Higgs 
model  in charm meson decays. We focus on the decays  
$D^+ \to \pi^+ l^+ l^-$ and  $D^0 \to \rho^0 l^+ l^-$, which  
are the most suitable 
  for the  experimental studies among all 
$D \to X l^+ l^-$ decay modes and they have the most stringent 
upper-bounds on the rates at present \cite{PDG}. We show that 
 the effects of LH model \cite{lee} can  slightly modify 
 the dilepton mass distribution  
for the inclusive decay $c \to u  l^+ l^-$. This effect is 
screened by the  long distance contributions in the hadronic channels. 
The total rate and the dilepton mass distribution for the decays 
$D^+ \to \pi^+ l^+ l^-$ and  $D^0 \to \rho^0 l^+ l^-$ are found to be 
completely dominated by the standard model long distance contributions. 
For this reason  we also 
reexamine the long distance contributions to $D^+ \to \pi^+ l^+ l^-$ 
using the most recent experimental results on charm meson decays.
We also consider the forward-backward asymmetry for the decay $D^0 \to \rho^0 l^+ l^-$, which is equal to zero in SM.  The forward-backward asymmetry is different from zero in LH model, but it 
is not large enough to be seen in the present or planned experiments.  
  
The paper is organized as follows. 
 In Section 2 we present the main results of the 
LH model by Lee \cite{lee}. In Section 3 we discuss the influence of 
this model on $c\to ul^+l^-$ transition. Section 4 and 5 are
 devoted to effects of LH model on decays $D^+ \to \pi^+ l^+ l^-$ and 
$D^0 \to \rho^0 l^+ l^-$, respectively. Our results are summarized
 in Section 6.

\section{FCNC in the Littlest Higgs model}

The Littlest Higgs models \cite{arkani} - \cite{csaki} offer an 
interesting and rather simple solutions to the  gauge hierarchy problem. 
 These models contain new massive gauge bosons and  a new heavy 
up-like quark $\tilde t$ together with its conjugate $\tilde t^c$. 
This quark  is a singlet under $SU(2)_{L}$, triplet under $SU(3)_{color}$ 
 and carries charge $2/3$ \cite{lee}. 
Its presence  modifies the weak currents \cite{lee}. 
The charged currents have SM contributions 
from the $W$ boson  as well as  the new contributions 
from a new gauge  boson $W_H$. 
The  SM Cabibbo-Kobayashi-Maskawa (CKM) matrix is 
extended to a $4 \times 3$ matrix.

 The neutral-current interactions in the SM do not change 
flavor at the tree level due to the GIM mechanism. 
The FCNC appears at one-loop level as a result of GIM cancellation and the 
difference of the quark masses. However, in 
the Littlest Higgs model the neutral current interactions  change the 
flavor already at the tree level. The Lagrangian which describes this 
interaction within the LH model is 
given by \cite{lee} 
\begin{equation}
{\cal L}_{NC} = \frac{g}{\cos \theta_W} Z_\mu (J_{W^3}^\mu - 
\sin^2 \theta_W J_{EM}^\mu ),
\label{e1}
\end{equation}
where $J_{EM}^\mu$ is the same electromagnetic current as in the SM, while $J_{W^3}^\mu$
is given by \cite{lee}
\begin{equation}
J_{W^3}^\mu = \frac{1}{2} \bar U_L^m \gamma^\mu \Omega U_L^m -  \frac{1}{2} \bar D_L^m \gamma^\mu  D_L^m
\label{e2}
\end{equation}
with $L=\tfrac{1}{2}(1- \gamma_5)$ and mass eigenstates $U_L^m= (u_L,c_L,t_L,T_L)^T$, $D_L^m=(d_L,s_L,b_L)^T$.
The neutral current for the down-like quarks is the same as in 
the SM,  
 while the up sector has additional currents since $\Omega\not = I$ due to the new 
 heavy quark \cite{lee}
\begin{equation}
\Omega = 
\begin{pmatrix}
1-|\Theta_u|^2 & - \Theta_u  \Theta_c^* & 
- \Theta_u  \Theta_t^* & - \Theta_u  \Theta_T^* \\
 - \Theta_c  \Theta_u^* & 1-|\Theta_c|^2  & 
- \Theta_c \Theta_t^* & - \Theta_c \Theta_T^* \\
- \Theta_t \Theta_u^* & - \Theta_t \Theta_c^* & 
1-|\Theta_t|^2  & - \Theta_t \Theta_T^* \\
-\Theta_T \Theta_u^* & - \Theta_T \Theta_c^* & 
- \Theta_T  \Theta_t^* &  1-|\Theta_T|^2  \\
\end{pmatrix}. \label{e3}
\end{equation} 
The elements of $\Omega$ satisfy following unitarity relations \cite{lee}:
\begin{align}
\label{e4}
|V_{id}|^2 + |V_{is}|^2 + |V_{ib}|^2+|\Theta_i|^2 & = 1~,\quad \ \ \ i = u,c,t,T\\
\label{e4a}
V_{id}V_{jd}^*+V_{is}V_{js}^*+V_{ib}V_{jb}^*+\Theta_i\Theta_j^*&=0~,\quad i,j = u,c,t,T ~,
\quad i\not = j~,
\end{align}
where $V_{ij}$ are CKM matrix elements. 
 The SM unitarity triangle is replaced by a 
unitary quadrangle in the LH Model. 

There is a tree-level flavor changing neutral coupling 
$\bar u_L\gamma_\mu c_LZ^\mu$ given by $ig\Omega_{uc}/(2\cos \theta_W)$ 
and we explore its possible effect in  rare charm meson decays. 
The magnitude of this  
effect depends on the value of $\Omega_{uc}=-\Theta_u \Theta_c^*$, 
 which is constrained  by the unitarity of CKM matrix via (\ref{e4}). 
  A more stringent upper bound 
on $\Omega_{uc}$ follows from the equality derived within LH model  \cite{lee}
\begin{equation}
\label{omega_uc}
 |\Omega_{uc}|\simeq |V_{ub}||V_{cb}|\frac{v^2}{f^2}\simeq 10^{-5}\biggl(\frac{1\ {\rm TeV}}{f}\biggr)^2~,
\end{equation} 
since the scale $f$ can not be arbitrarily small. 
At present, the scale $f$ is already severely constrained by the precision electro-weak observables. The lowest bound on $f$ 
ranges between $1$ TeV to $4$ TeV 
 or even higher, depending on the specific model \cite{csaki}, and  
we will vary the scale between
\begin{equation}
0.5\ {\rm TeV}\leq f\leq 4\ {\rm TeV}~. 
\end{equation} 
The scales below $1$ TeV are already excluded by the precision 
electro-weak observables \cite{csaki}, but we use them in order to demonstrate that even for the scale as low as $f=0.5$ TeV the effect of LH model on the rare charm meson decays is insignificant.

\section{Effects on  $c\to ul^+l^-$ transition}

The tree-level coupling $\bar u_L\gamma_\mu  c_LZ^\mu$ in LH model 
introduces new contributions in the effective weak Lagrangian 
relevant for $c\to ul^+l^-$ decay. 
Here we write only contributions which are relevant 
for our further study of charm meson 
decays\footnote{The notation for $C_{7,9,10}$ and $Q_{7,9,10}$ follows one 
given in Ref. \cite{jure}.}:
\begin{equation}
\label{lagrangian}
{\cal L}_{eff} = -\frac{G_F}{{\sqrt 2}} [ V_{cd}^*V_{ud}\sum_{i=1,2}C_iQ_i^d+V_{cs}^*V_{us}\sum_{i=1,2}C_iQ_i^s-V_{cb}^*V_{ub}\sum_{i= 7,9,10} C_i Q_i]~,
\quad 
\end{equation}
where quark operators are 
\begin{align}
\label{operators}
Q_9& = \frac{e^2}{16 \pi^2} \bar u_L \gamma_\mu c_L \bar l\gamma^\mu l ~,
\ \  Q_{10} = \frac{e^2}{16 \pi^2} \bar u_L \gamma_\mu c_L \bar l \gamma^\mu \gamma_5l~,\ \  Q_7=\frac{e}{8\pi^2}m_cF_{\mu\nu}\bar u\sigma^{\mu\nu}(1+\gamma_5)c~,\nonumber\\
Q_1^q&=\bar q_L \gamma^\mu q_L~\bar u_L\gamma_\mu c_L~,\quad\ Q_2^q=\bar u_L \gamma^\mu q_L~\bar q_L\gamma_\mu c_L~.
\end{align}

The SM  calculation leads to the prediction \cite{jure}\footnote{The branching ratio is expressed in terms of coefficients $C_i$ in \cite{prelovsek3,jure}. We use $m_c=1.4$ GeV.} 
\begin{equation}
\label{cull_sm}
\frac{\Gamma^{SM}(c\to ue^+e^-)}{\Gamma_{D^+}}=6.0\times 10^{-10}~,\quad 
\frac{\Gamma^{SM}(c\to u\mu^+\mu^-)}{\Gamma_{D^+}}= 1.3\times 10^{-10}~.
\end{equation}
Let us briefly describe  the dominant contributions that lead to this rate,  since we will  need the SM values of coefficients 
$C_{7,9,10}$ in the following sections.   The SM  rate 
is dominated by the photon exchange, 
where $c\to u\gamma$ is a two-loop diagram induced by $Q_2$  
and a gluon 
exchange \cite{greub,jure}. The  corresponding dominant piece in the amplitude
 is given by the 
coefficient 
$V_{cb}^*V_{ub}\hat C^{eff}_7=V_{cs}^*V_{us}(0.007+ 0.020i)(1\pm 0.2)$ 
\cite{greub,jure}\footnote{$\hat C^{eff}_7$ and $\hat C^{eff}_9$ are 
effective Wilson 
coefficients \cite{greub,jure}.} and tree-level matrix element 
$\langle Q_7\rangle_0$. The contribution of $\hat C_9^{eff}$, given by 
Eq. (7) of 
\cite{jure}, is small since it was found to be  significantly suppressed by 
the  effects of 
the renormalization group equations  for the  Wilson coefficients. 
The coefficient $C_{10}\simeq 0$ is completely 
negligible in the SM in contrast to the LH model, where it has the same 
magnitude as $C_9$ (\ref{C_LH}). 

\begin{figure}[htb!]
\begin{center}
\epsfig{file=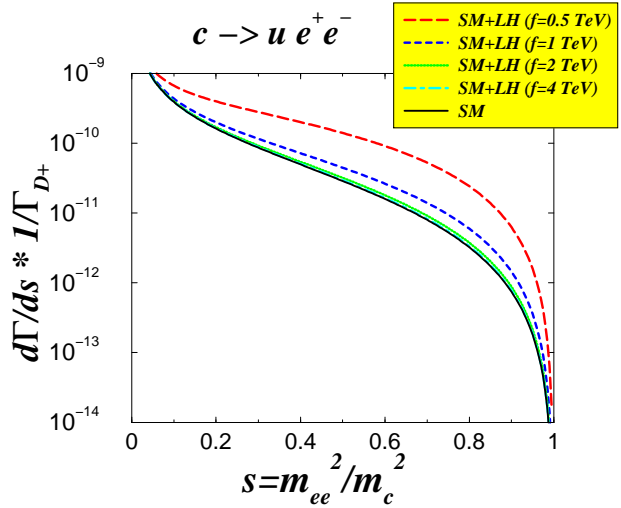,height=7cm}
\end{center}

\vspace{-0.8cm}

\caption{ \small The dilepton mass distribution  for the decay width of 
$c\to ue^+e^-$. The SM prediction \cite{jure} is shown together with 
possible modifications within LH model for various scales $f$.}
\label{fig.cull}
\end{figure}

The LH model contains the tree-level coupling $\bar u_L\gamma_\mu c_LZ^\mu$ (\ref{e1}) and modifies coefficients $C_9$ and $C_{10}$ 
\begin{equation} 
\label{C_LH}
 V_{cb}^* V_{ub}~ \delta C_{9}^{LH} = \frac{8\pi}{\alpha} \Omega_{uc} g_V^l,\quad
 V_{cb}^* V_{ub}~ \delta C_{10}^{LH} = -\frac{ 8\pi}{\alpha} \Omega_{uc} g_A^l
\end{equation}
with $g_V^l=-1/2+ 2 \sin^2 \theta_W$ and $g_A = -1/2$.  This model can 
moderately enhance the rate for the inclusive decay $c\to ul^+l^-$,  
 as illustrated for various scales $f$ in  
Figure \ref{fig.cull} \footnote{The phase of $\Omega_{uc}$ (\ref{omega_uc}) is unknown and we take the value that maximizes the rates in Figures \ref{fig.cull} and \ref{fig.pll}.}. 
The enhancement over the SM rate is practically 
negligible for the scales $f$ of few TeV or more. The enhancement is appreciable for $f\simeq 0.5~$TeV and we explore whether this could lead to 
any modifications of hadron observables in the following two sections. 

\section{Effects on  $D^+\to \pi^+ l^+l^-$ decay}

 The possible modification of $c\to ul^+l^-$ rates due to new physics can 
be probed experimentally only in the hadronic decays. 
First we focus on the $D^+\to \pi^+l^+l^-$ ($l=e,\mu$) decay, which  has the most stringent 
experimental upper bound 
 and  is the most promising for future experimental investigations among 
all $D\to X l^+l^-$ decays.  
The present experimental upper bounds are \cite{CLEO_pll,FERMILAB_pll,PDG}  
\begin{equation}
Br^{exp}(D^+\to \pi^+ e^+e^-)< 7.4\times 10^{-6}~,\quad 
Br^{exp}(D^+\to \pi^+ \mu^+\mu^-)< 8.8\times 10^{-6}~.
\end{equation}
The first rate concerns $m_{ee}$ outside the narrow region near 
$m_{ee}\simeq m_\phi$, while two events have already been observed in 
 the region where dilepton mass is close to the mass of $\phi$ meson 
 $m_{ee}\simeq m_\phi$ \cite{CLEO_pll} giving
\begin{equation}
\label{cleo2}
Br^{exp}(D^+\to \pi^+ \phi\to\pi^+e^+e^-)=(2.8\pm 1.9\pm 0.2)\times 10^{-6}~,
\end{equation}
which is consistent with 
$Br(D^+\to \phi \pi^+\to \pi^+e^+e^-)=Br(D^+\to \phi\pi^+)\times Br(\phi\to e^+e^-)=(1.9\pm 0.2)\times 10^{-6}$ \cite{PDG}.  

This indicates that the resonant decay channels 
$D^+\to \pi^+V_0\to \pi^+l^+l^-$  with 
intermediate vector resonances $V_0=\rho^0,\omega,\phi$ constitute 
an important long-distance contribution to the hadronic decay, 
which may shadow interesting 
short-distance contribution induced by $c\to ul^+l^-$ transition. Our 
determination of short and long distance contributions to 
$D^+\to \pi^+\l^+l^-$ takes advantage of the available experimental data. 
This is a fortunate circumstance for this particular decay since the 
analogous experimental input is not available for determination of  the other 
$D\to X l^+l^-$ rates in a similar way.  

\begin{figure}[htb!]
\begin{center}
\epsfig{file=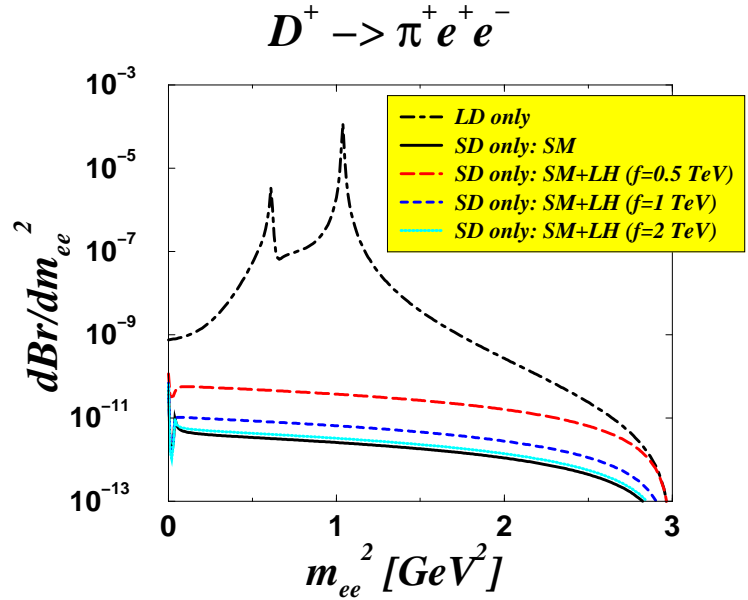,height=8cm}
\end{center}
\vspace{-0.8cm}

\caption{ \small The dilepton mass distribution 
 $dBr/dm_{ee}^2$ for the decay 
$D^+\to \pi^+e^+e^-$   
as a function of the  dilepton mass square $m_{ee}^2=(p_++p_-)^2$. 
 In the Standard and Littlest Higgs models the rate 
 is largely dominated by the 
long distance  contribution (LD), shown by the dot-dashed line. 
The other lines represent the short distance  contribution (SD) in SM and its 
modifications in LH model for various scales $f$.}\label{fig.pll}
\end{figure}

The size of the  short-distance contribution is dictated by the 
the coefficients $C_{7,9,10}$ in SM or LH model via
\begin{align}
\label{asd}
{\cal A}^{SD}[D(p)\to
\pi(p-q)l^+l^-]=i~\tfrac{G_F}{\sqrt{2}}e^2V_{cb}^*V_{ub}\biggl[\frac{C_{10}}{16\pi^2}f_+(q^2)~
&\bar u(p_-)\!{\not \!p}\gamma_5v(p_+)\\
+~\bigl\{\frac{C_7}{2\pi^2}m_c s(q^2)+\frac{C_9}{16\pi^2}f_+(q^2)\bigr\}
&\bar u(p_-)\!{\not \!p}v(p_+)~\biggr],\nonumber
\end{align}
where $q^2=m_{ll}^2$ and form factors $f_+(q^2)$ and $s(q^2)$ are defined by  
\begin{align}
\label{formsd}
 \langle \pi(p_\pi)|\bar
u\gamma^\mu(1-\gamma_5)c|D(p)\rangle&=(p+p_\pi)^\mu f_+(q^2)+(p-p_\pi)^\mu f_-(q^2)~,\\
\langle \pi(p_\pi)|\bar
u\sigma^{\mu\nu}(1\pm\gamma_5)c|D(p)\rangle&=
is(q^2)\bigl[(p+p_\pi)^\mu q^\nu-q^\mu(p+p_\pi)^\nu\pm i \epsilon^{\mu\nu\lambda\sigma}(p+p_\pi)_\lambda q_\sigma\bigr]\nonumber~.
\end{align}
We apply the information coming from recently measured 
decay distribution for $D\to\pi$ semileptonic decay \cite{fplus_bes,fplus}. 
They lead to the $D\to\pi$ form factor  
$f_+(q^2)=f_+(0)/(1-q^2/m_{D*}^2)$ with $f_+(0)=0.73\pm 0.14\pm 0.06$ \cite{fplus_bes}, which is consistent with $D^*$-pole dominance at present 
experimental accuracy \cite{fplus}. The experimental data for the form factor $s(q^2)$ is not 
available and we use the relation $s(q^2)=f_+(q^2)/m_D$ \cite{IW}, which strictly holds in 
the heavy quark limit and at zero recoil. 
The amplitude (\ref{asd}) gives the rates for the short distance contribution in Standard and Littlest Higgs models by using the values of coefficients $C_{7,9,10}$ from the previous Section. The resulting rates in 
Table \ref{tab.hadronic} and  Figure \ref{fig.pll}  indicate  that LH model can moderately enhance the short distance contribution in 
comparison with SM result.

\begin{table}[h]
\begin{center}
\begin{tabular}{|c|c|c|c|c|c|c|}
\hline
 {\bf Br} & \multicolumn{2}{c|}{short distance } & total rate $\simeq$ & experiment\\
 & \multicolumn{2}{c|}{contribution only } & long distance contr. & \\  

\hline 
 & SM & SM + LH  &    &   \\
& &         ($f=0.5~$TeV)  &    &   \\
\hline
$D^+\to \pi^+ e^+e^-$ & $6\times 10^{-12}$ & $8\times 10^{-11}$ & $1.9\times 10^{-6}$ & $<7.4\times 10^{-6}$\\
$D^+\to \pi^+ \mu^+\mu^-$ &  $6\times 10^{-12}$ & $8\times 10^{-11}$ & $1.9\times 10^{-6}$ & $<8.8\times 10^{-6}$\\
\hline
$D^0\to \rho^0 e^+e^-$ & negligible &$5\times 10^{-12}$ & $1.6\times 10^{-7}$ & $<1.0\times 10^{-4}$\\
$D^0\to \rho^0 \mu^+\mu^-$ & negligible &$5\times 10^{-12}$ & $1.5\times 10^{-7}$ & $<2.2\times 10^{-5}$\\
\hline
\end{tabular}
\caption{\small Branching ratios for the hadronic decays, which  are most suitable to probe $c\to ul^+l^-$ transition experimentally.  The total rates in Standard and Littlest Higgs models 
are completely dominated by the resonant 
long-distance contribution $D\to XV_0\to Xl^+l^-$.  We also provide the short-distance contribution in SM  together with its maximal modification in LH model for the scale  $f=0.5$ TeV. The SM short distance contribution for $D^0\to \rho^0l^+l^-$ is not shown since it is completely negligible in comparison to the long distance contribution. }\label{tab.hadronic}
\end{center}
\end{table} 

The  long-distance contributions arise from the  $D^+\to \pi^+V^0$ decay 
followed by $V_0\to \gamma\to e^+e^-$ where  $V_0=\rho^0,\omega~\phi$. 
The theoretical model for the long (and also short) 
distance contributions to all $D\to Pl^+l^-$ decays has 
been presented in \cite{prelovsek3}. 
The gauge invariance and Lorentz symmetry prohibit the decay  $D\to \pi \gamma$ to a real photon and there  is no $1/m_{ll}^2$ pole in the $D\to Pl^+l^-$ 
amplitude \cite{prelovsek0,prelovsek3}.   
Instead of using the theoretical model \cite{prelovsek3}, we  take the full advantage of  experimental input that is available for the decay of interest here. Our estimation is based on the measured rates for $D^+\to \pi^+\rho^0$, $\rho^0\to l^+l^-$,  $D^+\to \pi^+\phi$, $\phi\to l^+l^-$ \cite{PDG}
and the fact that the decay width for a cascade $D\to \pi V_0$ followed by $V_0\to l^+l^-$ 
can be generally expressed as \cite{lichard}\footnote{Relation (\ref{lichard}) applies for scalar resonances and also for vector resonances since $\bar l\! {\not\! q} l=0$ \cite{lichard}.}
\begin{equation}
\label{lichard}
\frac{d\Gamma_{D\to \pi V_0\to \pi l^+l^-}}{dq^2}=\Gamma_{D\to \pi V_0}(q^2)
\frac{1}{\pi}\frac{\sqrt{q^2}}{(m_{V_0}^2-q^2)^2+m_{V_0}^2\Gamma_{V_0}^2}\Gamma_{V_0\to l^+l^-}(q^2)~.
\end{equation}
 Here $\Gamma_{D\to \pi V_0}(q^2)$ and $\Gamma_{V_0\to l^+l^-}(q^2)$ denote 
rates if $V_0$ had a mass $\sqrt{q^2}$ and these rates are known experimentally only at  $\sqrt{q^2}=m_{V_0}$. Since vector resonances are 
relatively narrow, the relation 
(\ref{lichard}) can be further simplified using the narrow width approximation $\Gamma_{V_0}\ll M_{V_0}$ leading to
\begin{equation}
\label{product}
Br(D\to \pi V_0\to \pi l^+l^-)=Br(D\to \pi V_0)Br(V_0\to l^+l^-)~,
\end{equation}
which is in agreement with the experimental result (\ref{cleo2}). 
This indicates that 
the amplitude for a cascade via resonance $\rho^0$ or $\phi$ 
can be written as
\begin{equation}
{\cal A}^{LD}\bigl[D(p)\to \pi V_0\to \pi(p-q) l^+l^-\bigr]=e^{i\varphi_{\!\!\!~_{V_0}}}a_{V_0}~\frac{1}{q^2-m_{V_0}^2+i m_{V_0}\Gamma_{V_0}}~ \bar u(p_-){\!\not{\! p}} v(p_+)~,
\end{equation}
where the values $a_\rho=2.9\times 10^{-9}~$GeV$^{-2}$  and $a_\phi=4.2\times 10^{-9}~$GeV$^{-2}$
are determined from experimental data \cite{PDG}  via (\ref{product}) and the only 
assumption here is that  $a_{V_0}$ does not depend on $q^2$. Since the amplitude 
can be determined up to 
the overall  phase $e^{i\varphi_{\!\!\!~_{V_0}}}$ we keep it in our 
expressions. The contribution of the cascade via $\omega$ can not be determined in  such a   way since 
  only the upper limit on the  $D^+\to \pi^+\omega$ rate is  experimentally known. 

The long-distance amplitude is a sum of amplitudes for separate resonant channels, but their 
relative sign is not known since only the absolute value of $a_{V_0}$ can be 
determined from  (\ref{product}). We will argue that the relative signs as well as the 
ratio of $\omega/\rho^0$ amplitudes    
can be determined by considering the mechanism of the cascade decays. 
Our result is  
\begin{align}
\label{ald}
&{\cal A}^{LD}[D(p)\to \pi(p-q) l^+l^-]\\
&=e^{i\varphi}\biggl[a_{\rho}\biggl(\frac{1}{q^2-m_{\rho}^2+i m_{\rho}\Gamma_{\rho}}-\frac{1}{3}~\frac{1}{q^2-m_{\omega}^2+i m_{\omega}\Gamma_{\omega}}\biggr)-a_\phi \frac{1}{q^2-m_{\phi}^2+i m_{\phi}\Gamma_{\phi}}\biggr]\bar u(p_-){\! \not{\! p}} v(p_+)~,\nonumber
\end{align} 
where the values $a_{\rho,\phi}$ are given above, while the overall phase $\varphi$ is unknown but it is irrelevant since the phase of $\Omega_{uc}$ (\ref{omega_uc}) in ${\cal A}^{SD}$ (\ref{asd}) is unknown as well. 

The relative signs and the ratio of $\omega/\rho^0$ amplitudes can be 
 derived  by considering the mechanism of the decay 
$D^+\to \pi^+V^0\to \pi^+l^+l^-$. Part of the difference 
between amplitudes which proceed via $\rho^0,\omega,\phi$ comes from the 
electromagnetic (EM) transition $V^0\to\gamma\to \l^+l^-$, 
which depends on the quark 
content of the mesons in the  EM   
current $e_u\bar uu+e_d\bar dd+e_s\bar ss\to 
\tfrac{1}{\sqrt{2}}\rho^0+\tfrac{1}{3\sqrt{2}}\omega-\tfrac{1}{3}\phi$. 
 The remaining part of the difference is due to the 
weak transition $D^+\to\pi^+ V^0$, which is induced by the 
operators $Q_{1,2}^{d,s}$ (\ref{operators}) 
and  can proceed via three ways within the 
factorization approximation: 
\begin{enumerate}
 \item The first possibility is due to the operator $V_{cd}^*V_{ud}Q_1^d+V_{cs}^*V_{us}Q_1^s\simeq 
 V_{cd}^*V_{ud} \bar u_L\gamma^\mu c_L\times $ $(\bar d_L\gamma_\mu d_L-\bar s_L \gamma_\mu s_L)$, 
which induces the $D^+\to\pi^+$ transition via the $\bar u_L\gamma^\mu c_L$ current and 
 produces $\rho^0,\omega,\phi$ due to the acting of  the 
 $\bar d\gamma_\mu d-\bar s \gamma_\mu s$ current. The 
 $\bar dd\sim -\tfrac{1}{\sqrt{2}}\rho^0+\tfrac{1}{\sqrt{2}}\omega$ current 
 renders $\rho^0$ and $\omega$ with the opposite phase, while their amplitudes for the EM transition 
 differ by factor $1/3$, so 
$A_1(\omega)/A_1(\rho^0)=-1/3$ for this mechanism in the limit of $SU(3)$ flavor symmetry. Along the same lines $A_1(\phi)/A_1(\rho^0)=-2/3$. 

\vspace{-0.2cm}

\item The operator $Q_2^d$ can  
 induce $D^+\to \rho^0$ or $D^+\to \omega$ transition 
via the $\bar d_L \gamma^\mu c_L$ current and produce $\pi^+$ via  
$\bar u_L \gamma^\mu d_L$. Since $\rho^0$ and $\omega$ arise from 
$\bar dd$ again, this mechanism gives the same ratio 
$A_2(\omega)/A_2(\rho^0)=-1/3$, while 
there is no intermediate $\phi$ in this case.

\vspace{-0.2cm}

\item The third possibility arises from  $D^+$  which is annihilated by the  
$\bar d_L \gamma^\mu c_L$ operator and $V^0\pi^+$ created by the  
$\bar u_L \gamma^\mu d_L$ operator. It was shown within a model of \cite{prelovsek3} 
that this gives rise  only to bremsstrahlung diagrams and that the total bremsstrahlung amplitude is equal to zero for $D\to Pl^+l^-$ decays. So the model of \cite{prelovsek3} indicates that the contribution 
from this  mechanism is small and will be neglected. 
\end{enumerate}
Since the ratio of $\omega/\rho^0$ amplitudes is equal for 
first two mechanisms, the ratio for the total amplitudes 
 employed in 
(\ref{ald}) is $A(\omega)/A(\rho^0)=-1/3$.  The magnitude 
 $|A(\phi)/A(\rho^0)|=a_\phi/a_\rho$ is taken from
 the experimental data as explained above, while the  relative sign 
between 
$A(\phi)$ and $A(\rho^0)$  in (\ref{ald}) is negative due to 
the first mechanism, which is the only one that allows $\phi$ 
intermediate state. This explains the structure of amplitude presented in 
Eq. (\ref{ald}).

We point out that the long distance amplitude can not be determined in such a way for most of the remaining $D\to P l^+l^-$ decays. This is due to the lack of experimental data or to the fact that $A_1(\omega,\phi)/A_1(\rho^0)$  for the first mechanism may be different from  $A_2(\omega,\phi)/A_2(\rho^0)$ for the second mechanism. 

The amplitudes (\ref{asd}) and (\ref{ald}) give 
decay distributions in Figure \ref{fig.pll}, while the corresponding 
total rates are given in Table \ref{tab.hadronic}.  
  The rates in Standard and Littlest Higgs models are dominated by the 
resonant long distance contribution over the entire kinematical region 
of $m_{ll}^2$. Although the LH model with scale as low as 
$f=0.5~$ TeV  would enhance the 
short distance contribution,  it would not affect appreciably 
the dilepton mass distribution for $D^+\to \pi^+ l^+l^-$.

\section{Effects on  {\bf $D^0\to \rho^0 l^+l^-$} decay}

The $c\to ul^+l^-$ transition could be in principle also probed in 
$D\to Vl^+l^-$ decays with a vector meson $V$ in the final state. 
In this section we explore possible effects of the LH model 
on the rate of $D^0\to \rho^0 l^+l^-$, which has most strict 
experimental upper bound at present \cite{CLEO_vll,FERMILAB_vll,PDG} 
(see Table \ref{tab.hadronic}) 
and best prospects for future investigations. 

The long-distance contribution  $D^0\to \rho^0 V_0\to \rho^0 l^+l^-$ 
($V_0=\rho^0,\omega,\phi$) 
 is induced by $V_{cd}^*V_{ud}Q_1^d+V_{cs}^*V_{us}Q_1^s$ (\ref{operators}).
 We are unable to determine  its amplitude using the measured rates for 
$D^0\to \rho^0 V_0$ since only  the rate of $D^0\to \rho^0 \phi$ is known 
experimentally. We are forced to use a model and we apply the approach of 
\cite{prelovsek0} (an improved version of \cite{prelovsek2}), which 
 was developed to describe all $D\to Vl^+l^-$ and $D\to V\gamma$ decays. 
It is an effective model with mesonic degrees of freedom (heavy and light,  pseudoscalar and vector) and  is based on the heavy quark and chiral 
symmetries. The matrix elements are evaluated  using the factorization approximation 
and they are invariant under EM gauge transformation 
by construction. We apply the model and the 
values of the parameters from  section 5.4 of \cite{prelovsek0} to evaluate 
 the matrix elements for long and short distance contributions of $D^0\to 
\rho^0 l^+l^-$. 

\begin{figure}[htb!]
\begin{center}
\epsfig{file=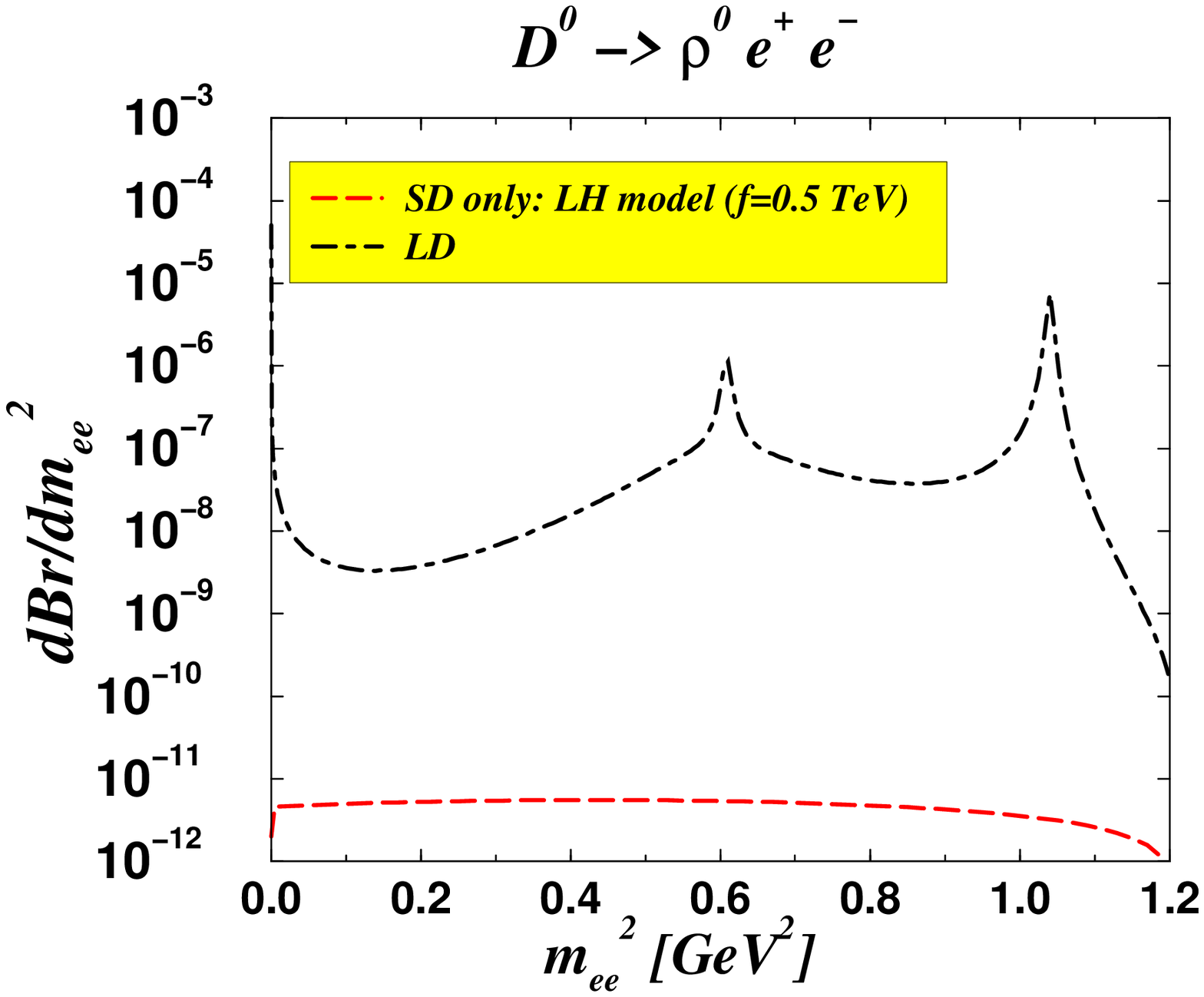,height=6cm} $\quad$
\epsfig{file=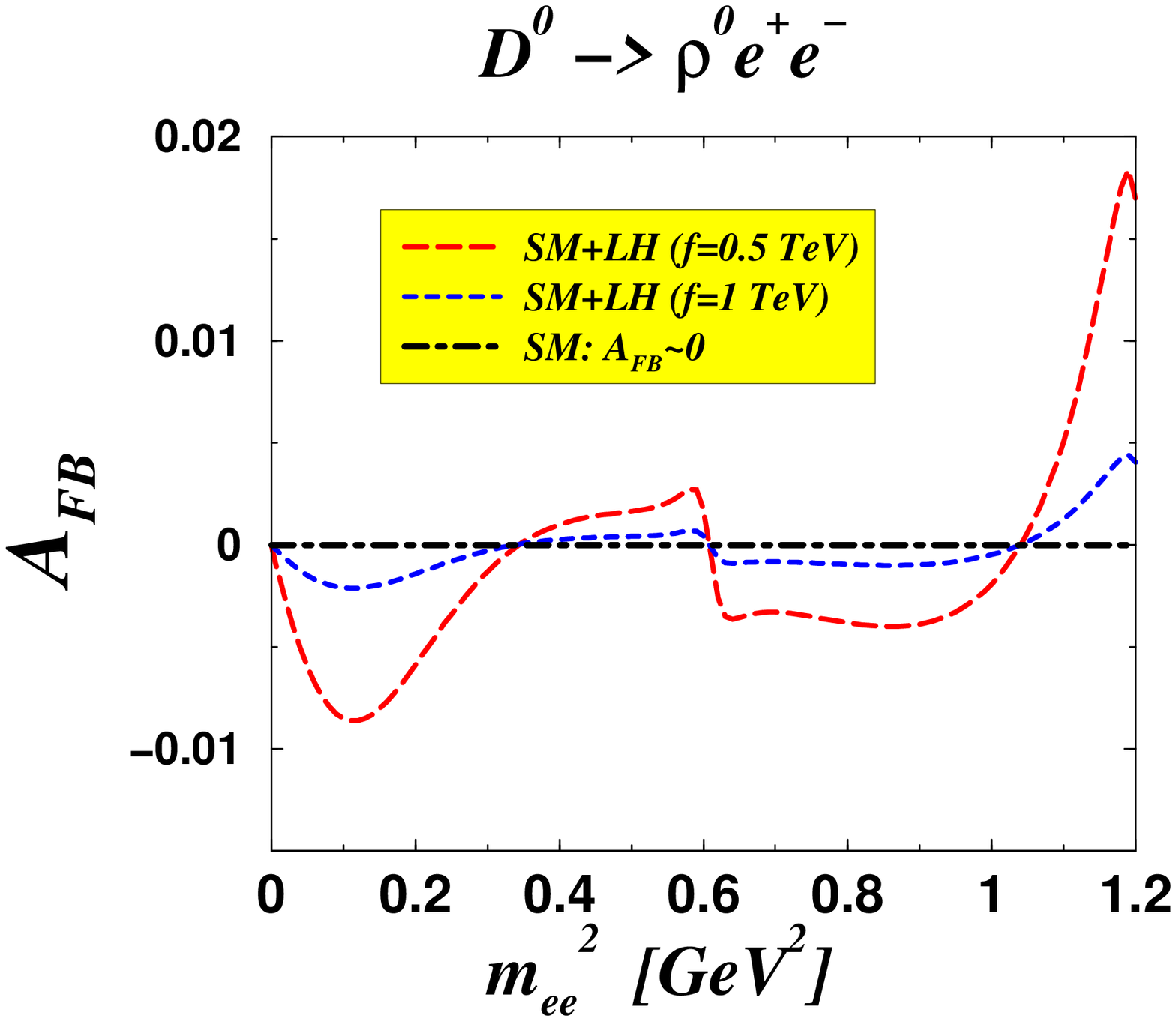,height=6cm}
\end{center}
\vspace{-0.8cm}

\caption{ \small  
The left figure shows the dilepton mass distribution for 
$D^0\to \rho^0e^+e^-$. It is completely dominated by the standard model long 
distance contributions (LD), shown by dot-dashed line. Dashed line illustrates the short distance contribution (SD) within LH model for $f=0.5$ TeV. The figure on the right shows the forward-backward asymmetry (\ref{afb}), which is zero in SM and small but nonzero in LH model.  }\label{fig.vll}
\end{figure}

The resulting long-distance contribution 
in Figure \ref{fig.vll} indicates that there 
is a pole at $m_{ll}\simeq 0$ in addition to the poles at $m_{ll}=m_{\rho,\omega,\phi}$. This 
pole is due to the photon propagator and arises 
since the decay $D^0\to \rho^0\gamma$ to a real photon is allowed\footnote{
The EM gauge invariance requires that $\langle \gamma |\bar d\gamma^\mu d-\bar s\gamma^\mu s|0\rangle 
\langle \rho^0 |\bar u_L\gamma_\mu c_L |D^0\rangle$ is zero 
for the  real photon in the factorization approximation \cite{soares}. The non-zero 
amplitude for $D^0\to \rho^0\gamma$ within the  
factorization approximation comes 
from the mechanism, where  $\bar u_L\gamma_\mu c_L$ annihilates $D^0$, 
$\bar d_L\gamma^\mu d_L$ creates $\rho^0$ and a photon is emitted before 
or after the weak transition.}. The long-distance contribution 
completely dominates the dilepton mass distributions 
in SM and  LH models. It also dominates the total rate given in Table \ref{tab.hadronic}. 
The short distance contributions are completely negligible  in SM  \cite{prelovsek0,prelovsek2} as well as in LH model even for the scale as low as $f=0.5~$ TeV. 

Our study shows that the LH model has negligible effect on the rate of $D^0\to \rho^0l^+l^-$, but it  might  have sizable effect  on
forward-backward asymmetry defined as 
\begin{equation}
\label{afb}
A_{FB}(m_{ll}^2)=\frac{\int_{0}^{1}\tfrac{d^2\Gamma}{d\cos\theta dm_{ll}^2}d\cos\theta-\int_{-1}^{0}\tfrac{d^2\Gamma}{d\cos\theta dm_{ll}^2}d\cos\theta}{\tfrac{d\Gamma}{dm_{ll}^2}}~,
\end{equation}
where $\theta$ is the angle between $l^+$ and $D^0$ in the $l^+l^-$ rest frame. We did not present $A_{FB}$ in the case of the $D\to \pi l^+l^-$ 
decay since it is 
equal to zero for the given amplitudes (\ref{asd},\ref{ald}).  
The non-zero asymmetry in $D\to \rho l^+l^-$ decay  arises 
only when $C_{10}\not =0$ (assuming $m_l\to 0$), 
so the asymmetry is practically zero in SM where $C_{10}\simeq 0$. 
The enhancement of the $C_{10}$ in the LH model (\ref{C_LH}) 
 is due to the tree-level $\bar u_L\gamma_\mu c_LZ^\mu$ coupling
 and leads to  nonzero asymmetry $A_{FB}(m_{ll}^2)$ shown in Figure \ref{fig.vll} \footnote{The asymmetry depends on the relative phase between short distance and long distance contributions and we fix this phase using $\arg(\Omega_{uc})=0$ in Figure \ref{fig.vll}.}. The asymmetry is of the order of $10^{-3}$ for $f\simeq 1$ TeV, but this is still too small to be observed in the present and foreseen experiments due to the smalness of the $D^0\to \rho^0l^+l^-$ rate.

\section{Conclusions}
The tree-level flavor changing neutral transition $c \to u Z$ appears 
within  a particular 
variation of the Littlest Higgs model \cite{lee}.  
 The magnitude of the relevant $c \to u Z$ coupling  
$|V_{ub}||V_{cb}|v^2/f^2$ 
is constrained via the scale $f\geq {\cal O}(1~{\rm TeV})$ 
 by the precision electro-weak data. 
We have investigated its impact   
on the rare $D$ meson decay observables.  First we determined 
the effects of the LH model on the  effective 
Wilson coefficients $C_9^{eff} $ and $C_{10}^{eff} $. 
Both coefficients have the same magnitude in the LH model, 
contrary to the result of SM 
where 
$C_{10}^{eff} \simeq 0$.  The LH model can appreciably modify the 
inclusive $c\to ul^+l^-$ decay only for the scales close to  $f=1$ TeV or 
less. 

Among  exclusive rare $D \to X l^+ l^-$ decays,  
the $D^+ \to \pi^+ l^+l^-$ and $D^0 \to \rho^0 l^+ l^-$ decays are the best 
candidates for the experimental searches  and have most stringent upper bounds at present. However, these decays are found to be completely dominated by the long distance contributions in SM as well as in LH models. Even the LH model with scale as low as $f=0.5$ TeV can not sizably modify the total rates and the dilepton mass distributions for $D^+ \to \pi^+ l^+l^-$ and $D^0 \to \rho^0 l^+ l^-$. The forward-backward asymmetry for  $D^0 \to \rho^0 l^+ l^-$ vanishes in SM, while it is of the order of $10^{-3}$ in LH model with the scale $f$ around $1$ TeV. Such asymmetry is still too small to be observed in the present or planned experiments given that the rate itself is already small.

We conclude that the LH model has insignificant effects on the charm meson observables in spite of 
its tree-level flavour changing couplings among up-like quarks. The eventual observation of the dilepton mass distribution and forward-backward asymmetry 
that disagree   with the standard model prediction would indicate the presence of some other scenario of physics beyond standard model.  

\vspace{1cm}

{\bf ACKNOWLEDGMENTS}

\vspace{0.3cm}

This research was supported in part by the Ministry of high education, science and technology of the 
Republic of Slovenia. One of us (SF) thanks A. Buras for mentioning  
the modifications in the  up-like sector within the LH model. She is very grateful for his warm hospitality during 
her Alexander von Humboldt stay at TU Munich, where this work has began.
We thank B. Bajc, T. Aliev, J. Zupan and J. Kamenik for many interesting discussions on this subject.   We also thank  J. Kamenik for reading the manuscript.

\end{document}